# Calculation of Elastic Strain Fields for Single Ge/Si Quantum Dots

Yu. N. Morokov*, M. P. Fedoruk

*Institute of Computational Technologies SB RAS, Novosibirsk, 630090 Russia and*
*Novosibirsk State University, Novosibirsk, 630090, Russia*

**Abstract** An atomistic model based on the Keating potential and the conjugate gradient method are used for simulation of the strain fields for single Ge/Si quantum dots. Calculations are performed in the cluster approximation using clusters containing about three million atoms belonging to 150 coordination spheres. The spatial distributions of the strain energy density and electron potential energy are calculated for different valleys forming the bottom of the silicon conduction band. It is shown that the strain field in silicon decreases sufficiently rapidly with distance from the center of the quantum dot, so the influence of the cluster boundary is observed only for very large quantum dots.

## I. INTRODUCTION

In previous work [1] it was shown that the development of multilayer structures with vertically stacked Ge/Si(001) quantum dots makes it possible to fabricate deep potential wells for electrons with vertical tunnel coupling. In this article it is given more detailed description of results of our calculations for strain fields for single Ge/Si quantum dots.

During the epitaxial growth of germanium on crystalline silicon, elastic strains develop at the heterointerface due to the different lattice constants of germanium and silicon. At the beginning of growth, a strained wetting layer is formed, which consists of several atomic germanium layers. In course of further growth, pyramidal islands are formed on the wetting layer's surface. The key role in the self-assembly of the formed nanostructures is played by inhomogeneous elastic strains in the system [2-4]. The formed islands several tens of nanometers in size are considered as quantum dots, since carriers localized in them exhibit a discrete spectrum.

As a rule, each study devoted to simulation of the electronic structure of self-assembled quantum dots starts with calculation of the elastic-strain distribution. This is due to the fact that the strain causes a change (distortion) in the band structure of the semiconductor materials; hence, has an effect on the shape of the potential well for electrons and holes in semiconductors. There are a lot of theoretical works on the strain distribution in quantum dot structures [4], including some works on modeling the electronic structure and strain distribution in self-assembled Ge/Si quantum dots [5-8].

In the Ge/Si heterostructure, holes are localized mostly in the bulk of quantum dots, in the germanium region. At the same time, the band offset makes the germanium region the potential barrier for electrons. Therefore, trapped electrons are mostly localized in silicon, and potential wells for them are formed due to the elastic deformation of silicon layers surrounding the quantum dots [6], near the tops of the quantum dots and under their bases. Due to their deformations, maximum splitting of the sixfold degenerate Δ-valley of the conduction band occurs in these regions. Two valleys, $\Delta^{001}$ and $\Delta^{00\bar{1}}$, oriented along the [001] and $[00\bar{1}]$ directions decrease in energy; the four others increase in energy. Localized electronic states are combined of the states of two bottom Δ-valleys.

In this paper, we calculate the spatial distributions of the strain energy density and electron potential energy near quantum dots for different valleys that form the bottom of the silicon conduction band.

## II. APPROACH DESCRIPTION

The discrete-continuous model with the Keating potential was used previously in [9] to simulate elastic-strain fields in quantum dots and in their surroundings. Some atoms in that model were considered in an explicit form; the effect of others was considered using Green's function calculated numerically. Unlike [9], in the present study and in [1] we use a more accurate atomistic model and cluster approximation. Initially, all atoms are arranged at sites of the perfect diamond-like silicon lattice with the constant $a_{Si}$ = 0.543072 nm which corresponds to the Si–Si bond length $l_{Si-Si}$ = 0.235137 nm. In this case, the distance between neighboring atomic layers in the [001] direction is $b_{Si} = a_{Si}/4$ = 0.135768 nm. Substitution of individual silicon atoms with germanium atoms in this lattice results in local stresses in the structure due to different equilibrium bond lengths Si–Si, Si–Ge, and Ge–Ge. It is assumed that the subsequent elastic relaxation in the system conserves the topology of interatomic bonds of the diamond-like structure.

We consider single quantum dots arranged on wetting layers five atomic layers thick. The considered quantum dots are {105} facetted pyramids with a square base and a height-to-base size ratio of 1 : 10. The calculated clusters are constructed by a sequential increase in the number of coordination shells, beginning from a certain central atom.

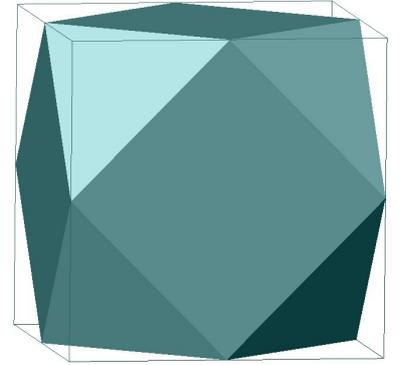

In picture the 14-hedron having a form of a cluster limited to the coordination sphere is presented. Such clusters were used by us in calculations. The polyhedron is obtained from a cube (along edges of this cube we direct the Cartesian axes of coordinates $x$, $y$, $z$) by cutting off eight vertexes of a cube on the main diagonals. As a result we obtain the 14-hedron having 6 square faces - the remains of initial faces of a cube, and 8 faces are the correct triangles as a result of cutting off 8 vertexes of a cube.

The number of atoms in the cluster, which is limited by the coordination sphere with number $n_c$, is determined by

$$n_a = \frac{5}{6}n_c^3 + \frac{5}{4}n_c^2 + \frac{13}{6}n_c + 1, \quad n_c = 0, 2, 4, \ldots,$$

$$n_a = \frac{5}{6}n_c^3 + \frac{5}{4}n_c^2 + \frac{13}{6}n_c + \frac{3}{4}, \quad n_c = 1, 3, 5, \ldots.$$

The number of atoms in the coordination sphere with number $n_c$ is equal

$$n_s = \frac{5}{2}n_c^2 + 2, \quad n_c = 2, 4, \ldots,$$

$$n_s = \frac{5}{2}n_c^2 + \frac{3}{2}, \quad n_c = 1, 3, 5, \ldots.$$

In the case of elastic relaxation of the system, the system's energy functional written as the Keating potential [10]

$$E = \frac{3}{16}\sum_i \sum_j \frac{\alpha_{ij}}{l_{ij}^2} \cdot \left[(\mathbf{r}_i - \mathbf{r}_j)^2 - l_{ij}^2\right]^2 + \frac{3}{8}\sum_i \sum_{j>k} \frac{\beta_{ijk}}{l_{ij} \cdot l_{ik}} \cdot \left[(\mathbf{r}_i - \mathbf{r}_j)\cdot(\mathbf{r}_i - \mathbf{r}_k) + \frac{1}{3}l_{ij} \cdot l_{ik}\right]^2$$

is minimized, where $\mathbf{r}_i$ is the radius vector of the $i$-th atom: $\alpha_{ij}$, $\beta_{ijk}$, and $l_{ij}$ are the parameters depending on the atom type (the subscript $i$ enumerates all cluster atoms, and subscripts $j$ and $k$ enumerate the nearest neighbors of the $i$-th atom). The parameters $\alpha_{ij}$ and $\beta_{ijk}$ play the role of force constants, and $l_{ij}$ are the equilibrium bond lengths between atoms.

In this work, also as in [1] and [9], values of parameters are taken from [10]: $\alpha$ = 48.5 (Si), 38 (Ge) N/m, $\beta$ = 13.8 (Si), 12 (Ge) N/m. Parameters $\alpha_{ij}$, $\beta_{ijk}$ and $l_{ij}$ for atoms of different types are taken as the arithmetical mean [9]: $l_{SiGe}$ = ($l_{SiSi}$ + $l_{GeGe}$)/2; $\alpha_{SiGe}$ = ($\alpha_{SiSi}$ + $\alpha_{GeGe}$)/2; $\beta_{ijk}$ = (($\beta_{iii}$ + $\beta_{jjj}$) + ($\beta_{iii}$ + $\beta_{kkk}$))/4.

The total energy of the system can be written as the sum of $E_i$ corresponding to contributions of individual atoms of the system,

$$E_i = \frac{3}{16}\sum_i \frac{\alpha_{ij}}{l_{ij}^2} \cdot \left[(\mathbf{r}_i - \mathbf{r}_j)^2 - l_{ij}^2\right]^2 + \frac{3}{8}\sum_{j>k} \frac{\beta_{ijk}}{l_{ij} \cdot l_{ik}} \cdot \left[(\mathbf{r}_i - \mathbf{r}_j)\cdot(\mathbf{r}_i - \mathbf{r}_k) + \frac{1}{3}l_{ij} \cdot l_{ik}\right]^2.$$

The quantity $E_i$ can be interpreted as the fraction of the elastic energy related to the $i$-th atom. The system energy is minimized using the conjugate gradient method. The numerical calculation is completed, when the change in the total cluster energy at one conjugate gradient step becomes smaller than the total cluster energy by 14 orders of magnitude. The accuracy limitation is caused by rounding errors when using double-precision numbers.

We use the following boundary conditions for the clusters. For atoms of two outer coordination shells (which also include germanium atoms of the wetting layer), the $x$- and $y$-coordinates are fixed, but the $z$-coordinates (in the growth direction) are completely free for the relaxation of all atoms. It becomes possible due to a rupture of crystal silicon by germanium of the wetting layer infinite in the x and y directions.

Basic calculations are performed for clusters containing atoms of 150 coordination shells; the central cluster atom is considered as belonging to the zeroth coordination sphere. The cluster of 150 coordination shells contains 2840951 atoms. Two last coordination spheres contain 111756 atoms, and, respectively, only atoms of 148 coordination spheres (2 729 195 atoms) are completely free for the relaxation.

To calculate the single-particle quantum electronic states localized near the quantum dot, the effective electron potential energy can be written as the sum of the potential energy without regard for the lattice strain and the potential related to the elastic strain. If we use the effective-mass approximation and take the bottom of the conduction band as the reference point of electron energy, the electron potential energy in the strained silicon is reduced to the energy related to the lattice elastic strain.

Generally speaking, the lattice strain removes valley degeneracy, and the electron energy for different valleys is given by [11]

for $\Delta^{100}$ and $\Delta^{\bar{1}00}$    $U_e = \Xi_d u(\mathbf{r}) + \Xi_u u_{xx}(\mathbf{r})$,

for $\Delta^{010}$ and $\Delta^{0\bar{1}0}$    $U_e = \Xi_d u(\mathbf{r}) + \Xi_u u_{yy}(\mathbf{r})$,

for $\Delta^{001}$ and $\Delta^{00\bar{1}}$    $U_e = \Xi_d u(\mathbf{r}) + \Xi_u u_{zz}(\mathbf{r})$.

Here, $\Xi_d$ and $\Xi_u$ are the strain potential constants, $u_{\alpha\beta}(\mathbf{r})$ is the strain tensor at the point $\mathbf{r}$, and $u(\mathbf{r})$ = $u_{xx}(\mathbf{r})$ + $u_{yy}(\mathbf{r})$ + $u_{zz}(\mathbf{r})$ is the strain tensor trace. We used the same strain potential constants for Si as in [9, 11]: $\Xi_d$ = 1.28 eV, $\Xi_u$ = 8.7 eV. We do not use the corresponding parameters for germanium because we calculate the electron potential energy only for the "volume" silicon atoms bonded only with other silicon atoms.

Below, by consideration of the electron potential energy, for example, for $\Delta^{100}$ valley we will automatically mean that this consideration applies also to $\Delta^{\bar{1}00}$ valley. Therefore valleys $\Delta^{\bar{1}00}$, $\Delta^{0\bar{1}0}$, and $\Delta^{00\bar{1}}$ will not be noted separately.

The following algorithm is used for calculation of the strain tensor [9].

The values of the components of the strain tensor are calculated only for the nodes (centers of atoms) of the crystal lattice. Let us place the origin of the Cartesian coordinate system (*x, y, z*) in the node, for which we will calculate the strain tensor. The centers of the four atoms closest to the node form the vertices of a 4-hedron. The four points with the radius vectors

$$\mathbf{r}^{(1)} = \pm \frac{d_1}{\sqrt{3}}(1,1,1), \quad \mathbf{r}^{(2)} = \pm \frac{d_2}{\sqrt{3}}(1,-1,-1), \quad \mathbf{r}^{(3)} = \pm \frac{d_3}{\sqrt{3}}(-1,1,-1), \text{ and } \mathbf{r}^{(4)} = \pm \frac{d_4}{\sqrt{3}}(-1,-1,1)$$

are considered as vertexes of tetrahedron of unstrained ideal lattice. The symbols $d_1$ - $d_4$ are the equilibrium bond lengths of the corresponding atomic bonds, the "+" or "-" is selected, depending on the sublattice. Let $\mathbf{r}'^{(1)}$, $\mathbf{r}'^{(2)}$, $\mathbf{r}'^{(3)}$, and $\mathbf{r}'^{(4)}$ designate the radius vectors of vertexes of deformed tetrahedron in the deformed lattice.

The displacement vectors $\mathbf{u}^{(i)} = \mathbf{r}'^{(i)} - \mathbf{r}^{(i)}$, $1 \le i \le 4$ are calculated. Nine quantities $w_{\alpha\beta}$ are determined by

$$w_{\alpha x} = \left(u_\alpha^{(1)} + u_\alpha^{(2)} - u_\alpha^{(3)} - u_\alpha^{(4)}\right)/a,$$
$$w_{\alpha y} = \left(u_\alpha^{(1)} + u_\alpha^{(3)} - u_\alpha^{(2)} - u_\alpha^{(4)}\right)/a,$$
$$w_{\alpha z} = \left(u_\alpha^{(1)} + u_\alpha^{(4)} - u_\alpha^{(2)} - u_\alpha^{(3)}\right)/a,$$

where *a* is a lattice constant. The quantities $w_{\alpha\beta}$ correspond to derivatives $\partial u_\alpha / \partial r_\beta$ in continuum approach. According to [12] the strain tensor $u_{\alpha\beta}$ is defined by

$$u_{\alpha\beta} = \frac{1}{2}\left(w_{\alpha\beta} + w_{\beta\alpha} + w_{\gamma\alpha}w_{\gamma\beta}\right).$$

This algorithm for estimating the values of the components of the strain tensor is simple and economical in realization. Although the algorithm is only a first order approximation in terms of continuum approach, but in our case, the discrete atomistic model is initially considered on the atomic scale. In such a situation, continuum approach itself is only a rough approximation to the discrete reality. Therefore, we can only speak of a reasonable estimating of the values of the components of the strain tensor.

### III. RESULTS

**1. Relaxation of the wetting layer**

Let us first consider the relaxation of the wetting layer without pyramids.

A free relaxation of *z*-coordinates of all atoms allows germanium atoms of the wetting layer partially reduce stress through relaxation along the *z*-direction. Results of calculations of local strain energy counting on one germanium atom and the diagonal components of the strain tensor are given in Table 1 for two wetting layers with thickness of 5 and 10 atomic layers.

As the wetting layers are relaxed within the Keating model without changing the topology of the interatomic bonds, we cannot estimate the stability of these layers with respect to the formation of island structures on them.

At relaxation of a cluster with the wetting layer, the *x*- and *y*-coordinates of atoms stay practically unchanged, and a relaxation is reduced only to change of their *z*-coordinates. Therefore, negative values of the components of the strain tensor $u_{xx} = u_{yy}$ correspond to the initial deformation of atoms (compression in the plane of the layer) and do not depend on thickness of the wetting layer.

| | 5 atomic layers | | | | 10 atomic layers | | |
|---|---|---|---|---|---|---|---|
| | Energy (meV/atom) | $u_{xx} = u_{yy}$ | $u_{zz}$ | | Energy (meV/atom) | $u_{xx} = u_{yy}$ | $u_{zz}$ |
| | | | 0.00012 | Si | 0.000 | 0.00000 | 0.00012 |
| Si | 0.008 | 0.00000 | -0.00126 | Si | 0.008 | 0.00000 | -0.00126 |
| Si | 4.209 | -0.02066 | 0.01413 | Si | 4.208 | -0.02066 | 0.01413 |
| Ge | 20.041 | -0.06156 | 0.04558 | Ge | 20.040 | -0.06155 | 0.04558 |
| Ge | 31.506 | -0.08178 | 0.05869 | Ge | 31.504 | -0.08178 | 0.05868 |
| Ge | 31.501 | -0.08178 | 0.05732 | Ge | 31.498 | -0.08178 | 0.05744 |
| | | | | Ge | 31.498 | -0.08178 | 0.05757 |
| | | | | Ge | 31.498 | -0.08178 | 0.05755 |

**Table 1.** The calculated values for the relaxed wetting layers consisting of 5 and 10 atomic layers of germanium atoms. The data are given for the atomic layers, which are located above the middle layer, including layers of germanium atoms and the three nearest silicon atomic layers.

The data in Table 1 show that the initial negative value -0.08178 for the $u_{zz}$ components of the strain tensor (compression along the axis $z$) for germanium atoms of the wetting layer is changed to positive (stretching along the axis $z$) in the relaxation process. Stretching thus becomes almost uniform in the wetting layer with $u_{zz} = 0.058$, independently of the thickness of the wetting layer. Transition from this value to $u_{zz} = 0$, corresponding non-strained silicon, occurs on the scale of a few atomic layers, regardless of the thickness of the wetting layer.

The sum of the diagonal components of the strain tensor $u_{xx} + u_{yy} + u_{zz}$ is the relative volume change for a small element. From Table 1 it is seen that the stretching of germanium atoms along the $z$ axis does not compensate their compression in $x$-$y$ plane, and the volume per one germanium atom remains considerably smaller after relaxation than in crystalline germanium.

In the following calculations we consider wetting layers consisting of only 5 atomic layers.

For the cluster containing atoms of 150 coordination spheres with one wetting layer thickness of 5 atomic layers of germanium, the total initial energy of the cluster (before relaxation) was equal 3738.572 eV. The total energy decreased to 1602.395 eV in the process of relaxation.

## 2. Single quantum dot with a base half-width of 55 atomic layers

This germanium pyramid surrounded with crystal silicon is shows in Fig. 1. We put the origin of the Cartesian coordinate system on the central atom of cluster (it is highlighted in Fig. 1). This atom is also the central atom in the square basis of the pyramid, i.e. in the first atomic layer over the wetting layer.

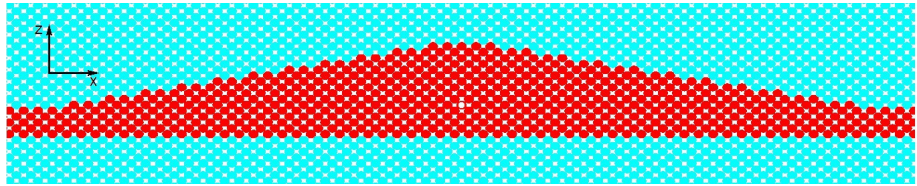

**Figure 1.** The central section ($y = 0$) of the single germanium pyramid surrounded with crystal silicon. The base of the pyramid consists of the atoms of 111 atomic layers along the $x$ axis.

More often, the central sections ($y = 0$) of the clusters will be presented in the subsequent figures. Cases of consideration of other, lateral sections ($z$ = const), will be separately mentioned.

The initial energy of the cluster was 4199.064 eV. The initial maximum value of components of the energy gradient (projections of forces on the Cartesian axes) was 47.9 eV/nm. The energy decreases to 1811.818 eV after 819 steps of the method of conjugate gradients, and 90% of energy reduction is achieved during the first 30 steps. A reduction of energy for one step was about $10^{-10}$ eV at final step, and the maximum value of components of the energy gradient was $2.85 \cdot 10^{-6}$ eV/nm.

## 2.1. The strain distribution in germanium for the single pyramid

The top part of Fig. 2 shows the full cross-section of the wetting layer and the pyramid. The width of the wetting layer along the $x$ axis is equal to the width of the cluster. The wetting layer contains atoms of 301 atomic layers (along the $x$ axis) and its width is equal 40.7304 nm (between the centers of the extreme atoms).

For more informative display of energy in the colors we use unlinear color scale. The concordance between energy values and the color palette is established in the right side of Fig. 2.

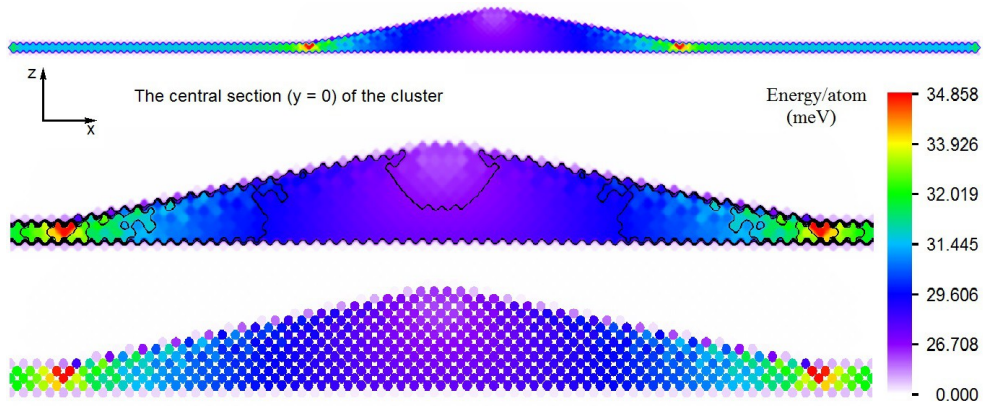

**Figure 2.** The distribution of the bulk density of the strain energy in germanium with a base half-width of 55 atomic layers.

From the top part of Fig. 2 it is visible that the strain energy density in the wetting layer decreases quickly enough to the value received for the wetting layer without pyramid (31.50 meV/atom, Table 1) with increasing the distance to the center of the pyramid. This demonstrates the adequacy of used cluster size (150 coordination spheres) to simulate the strain field in the vicinity of the pyramid.

The distribution of the residual strain energy in the pyramid and near its edges is represented in more detail in lower parts of Fig. 2. The middle fragment of Fig. 2 shows the contour lines corresponding to the numerical energy values given in the palette, and the lower fragment shows the strain energy density with atomic resolution.

The maximum value 34.858 meV/atom of the strain energy density in the cluster is achieved on the atoms of the wetting layer. These atoms are remote from outside atoms of the pyramid on distance of two atomic layers both along the $x$-axis and along the $z$-axis. Increasing the strain energy on 3.35 meV for germanium atoms of the wetting layer near the pyramid is connected with the pressure of the array of germanium atoms of the pyramid, which endeavor to reduce their compressive strain in the lateral direction.

Figure 3 shows the distribution of the strain energy density in the first lower layer of the pyramid atoms ($z = 0$). The maximum density of strain energy (32.410 meV/atom) directly for atoms of the pyramid is observed in this layer near the edge of the step formed by atoms of the second layer of a pyramid. The strain energy decreases to 26.873 meV/atom in the center of this layer (Table 2).

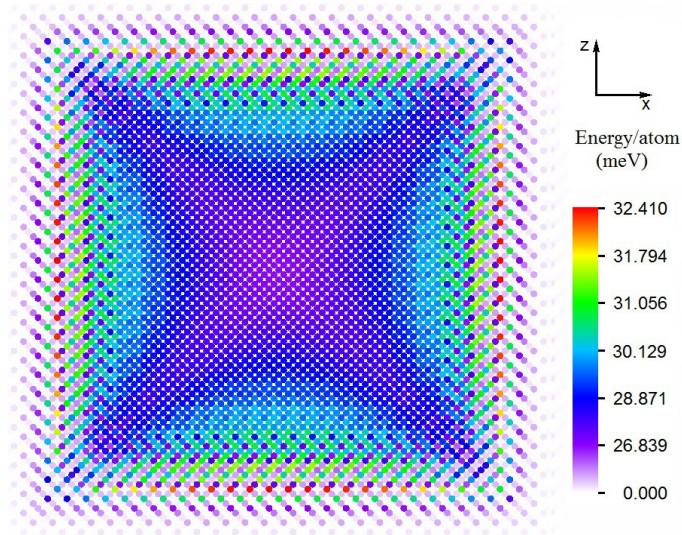

**Figure 3.** The distribution of the bulk density of the strain energy in germanium in the lateral section ($z = 0$).

The distributions of computed values along a pyramid axis (the axis $z$) are shown in Table 2. For calculation of these values we considered a vertical column directed along the pyramid axis with a square cross-section of 8 × 8 atomic layers. This cross-section coincides by the sizes with a plateau on top of this pyramid. The table shows the values obtained by averaging over the atoms of the column which have the same initial $z$-coordinates. The numbers of atoms in such cross-sections of the column are equal to 8, 13, 8 and 12 (with a period in four atomic layers along the axis $z$).

**Table 2.** The distributions of the bulk density of the strain energy and the strain tensor components. Positive layer numbers correspond to germanium atoms of the pyramid and two layers of silicon over a top of the pyramid. Negative layer numbers correspond to the wetting layer of germanium atoms and atoms of the next two layers of silicon.

| Layer number | Atoms | Energy (meV) | $u_{xx} = u_{yy}$ | $u_{zz}$ |
|---|---|---|---|---|
| 12 | Si | 2.426 | 0.01063 | -0.03216 |
| 11 | Si | 4.251 | -0.00803 | -0.02323 |
| 10 | Ge | 13.118 | -0.04264 | -0.00237 |
| 9 | Ge | 25.176 | -0.06988 | 0.01800 |
| 8 | Ge | 26.059 | -0.07158 | 0.02215 |
| 7 | Ge | 25.136 | -0.07068 | 0.02273 |
| 6 | Ge | 25.975 | -0.07239 | 0.02679 |
| 5 | Ge | 25.761 | -0.07226 | 0.02763 |
| 4 | Ge | 26.221 | -0.07315 | 0.02985 |
| 3 | Ge | 26.228 | -0.07330 | 0.03081 |
| 2 | Ge | 26.508 | -0.07385 | 0.03244 |
| 1 | Ge | 26.624 | -0.07413 | 0.03348 |
| 0 | Ge | 26.873 | -0.07459 | 0.03480 |
| -1 | Ge | 27.002 | -0.07485 | 0.03568 |
| -2 | Ge | 27.202 | -0.07521 | 0.03676 |
| -3 | Ge | 27.361 | -0.07549 | 0.03748 |
| -4 | Ge | 27.489 | -0.07584 | 0.03969 |
| -5 | Ge | 17.070 | -0.05577 | 0.02777 |
| -6 | Si | 3.596 | -0.01501 | -0.00208 |
| -7 | Si | 0.570 | 0.00555 | -0.01656 |

As seen from Table 2 and Fig. 2, a tendency to decrease the strain energy at the transition from the base of the pyramid to its apex (along an axis $z$) is observed for germanium atoms.

It is visible also from Fig. 3 that the use of the central cross-section ($y = 0$) presented on Fig. 2 and the subsequent figures, allows to get an idea about the maximum strains in the cluster.

Figures 4 and 5 show the distributions of $u_{xx}$ and $u_{yy}$ components of the strain tensor for the central section of the pyramid. These components are negative for all germanium atoms, meaning compression in *x-y* plane.

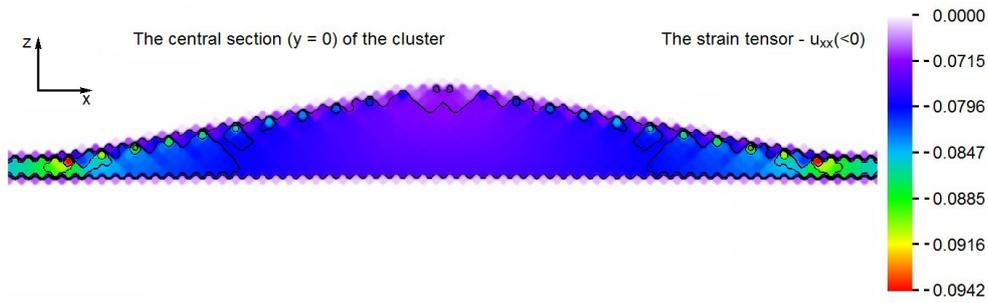

**Figure 4.** Distribution of the strain tensor components $u_{xx}$.

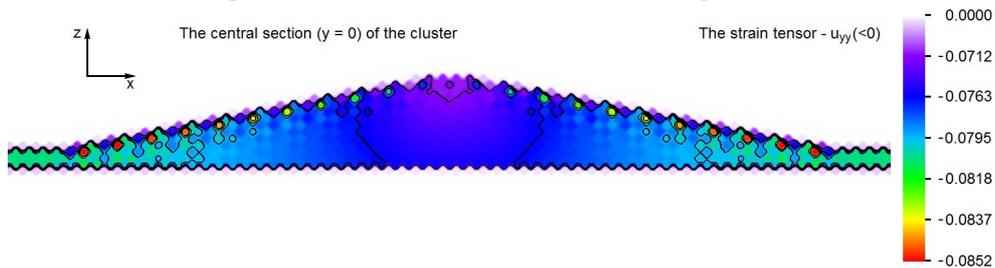

**Figure 5.** Distribution of the strain tensor components $u_{yy}$.

The maximum absolute value $u_{xx}$ = -0.0942 achieved on the same atoms of the wetting layer near the edge of the pyramid, where the maximum value of the strain energy density achieved. This value exceeds the absolute value of $u_{xx}$ = -0.0818 in the wetting layer without pyramid (Table 1). Compression in *x-y* plane is somewhat smaller in magnitude in the central region of the pyramid. Local centers of increased values of $u_{xx}$ on sloped faces of the pyramid (light areas) are located in the second coordination sphere in relation to the edge atoms of steps on these faces.

More detail distributions of $u_{xx} = u_{yy}$ and $u_{zz}$ along the pyramid axis are shown in Table 2.

Figure 6 shows the space distribution of only positive components $u_{zz}$ of the strain tensor (negative values are displayed in white). All germanium atoms are exposed of stretching in the direction of the axis *z* (*z* > 0), with the maximum degree of stretching near the side edge of the pyramid ($u_{zz}$ = 0.0703) that more than 3 times exceeds the stretching near the pyramid apex.

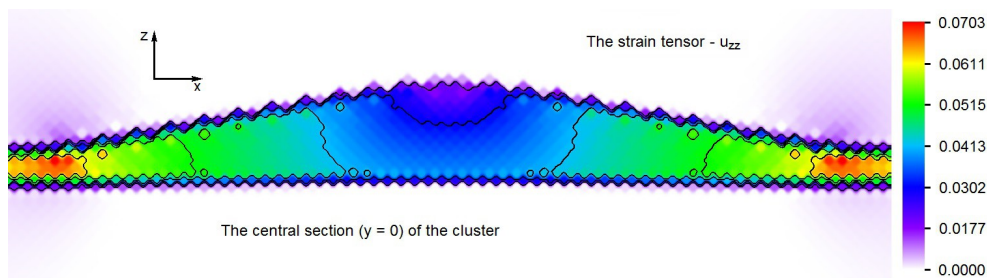

**Figure 6.** Distribution of the strain tensor components $u_{zz}$.

For all germanium atoms, stretching of along the *z* axis does not compensate for their compression in *x-y* plane, and the volume per one germanium atom remains after relaxation noticeably less than in crystalline germanium.

The components $u_{xx}$ and $u_{yy}$ are negative and the components $u_{zz}$ are positive in the lateral section of the first layer of the pyramid ($z = 0$). Figures of corresponding distributions visually actually coincide with Fig. 3 for the distribution of the strain energy density.

The germanium pyramid can capture holes. Trapped holes are mostly localized in the bulk of quantum dots, in the germanium region. The distribution of the deformation potential, which is determined by the strain tensor, has a significant impact on the parameters of the quantum states for holes. At the same time, trapped electrons are mostly localized in silicon, and potential wells for them are formed due to the elastic deformation in silicon surrounding the quantum dots.

### 2.2. The strain distribution in silicon for the single pyramid

Here we use results of the same calculation of the pyramid. But now we will use other nonlinear scale of a palette for visualization.

Figure 7 shows the distribution of the bulk density of the strain energy in silicon surrounding the pyramid.

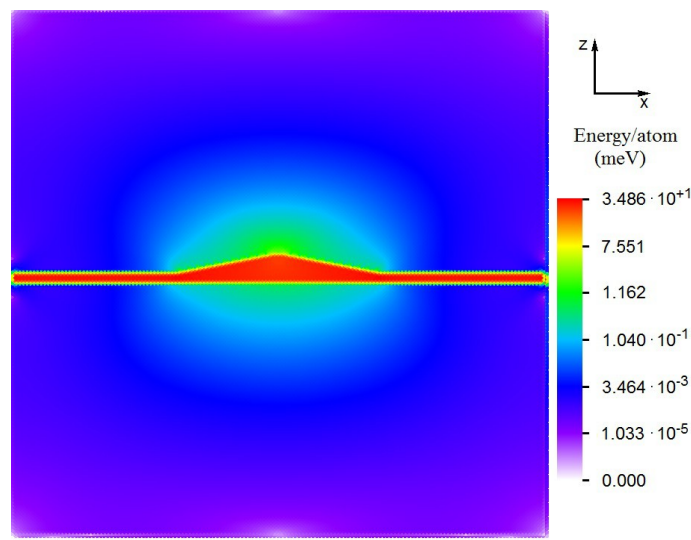

**Figure 7.** The distribution of the bulk density of the strain energy in silicon in the central section of the cluster containing a single quantum dot with a base half-width of 55 atomic layers.

Change of strain energy density in silicon along the pyramid axis is presented in Fig. 8. These values were obtained by averaging on cross-sections of $8 \times 8$ atomic layers of square column, also as well as the data in Table 2.

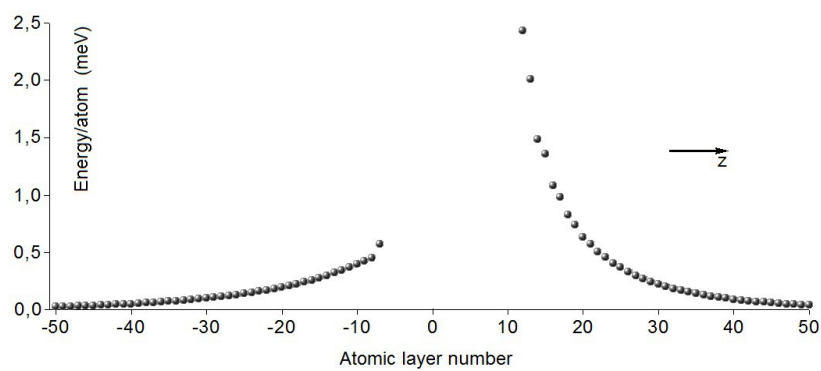

**Figure 8.** The distribution of the bulk density of the strain energy in silicon along the pyramid axis.

The graph does not display higher energies for germanium atoms and for the silicon atoms forming Si-Ge bonds. That is, the layers from -6 to 11 are not displayed. Values for these layers are given in Table 2.

It is visible from Fig. 7 and Fig. 8 that deformation in silicon decreases rather quickly with moving away from the pyramid. The strain energy in silicon in the nearest vicinity of the pyramid is larger near the pyramid apex than under its base.

The nondiagonal components $u_{xz}$ of the strain tensor are responsible for a shear deformation of the lattice in *x-z* plane. Positive values are observed in two quadrants of the plane (Fig. 9). The picture is similar in two other quadrants, but with a negative sign. The maximum shear displacements in this plane are observed near the two side faces of the pyramid.

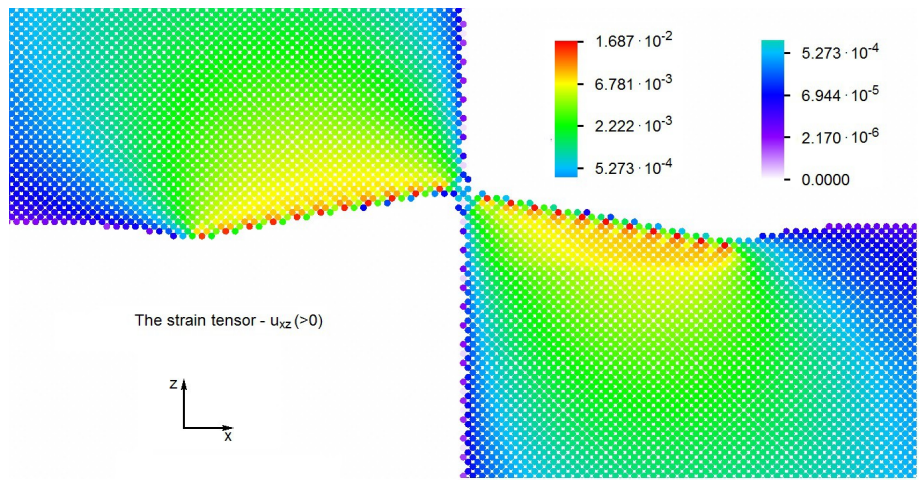

**Figure 9.** The distribution of the positive nondiagonal components $u_{xz}$ of the strain tensor.

## 2.3. The electron potential energy in silicon for $\Delta^{001}$ valley for the single pyramid

The distribution of the electron potential energy in silicon for $\Delta^{001}$ valley in the central section of the cluster is presented in Fig. 10.

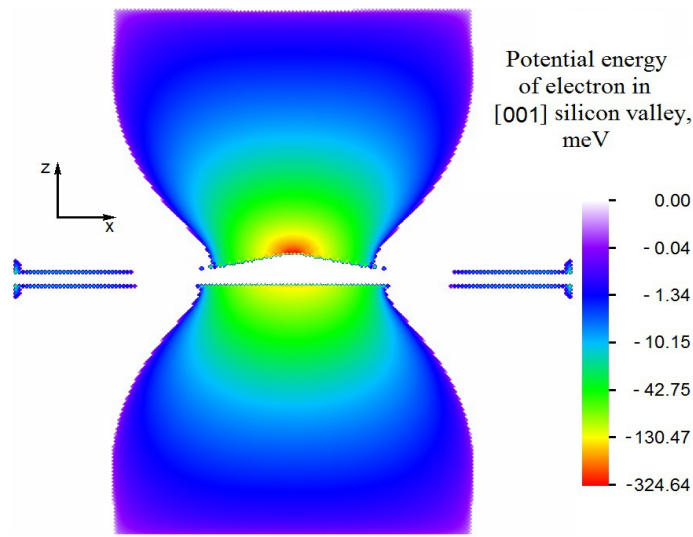

**Figure 10.** The distribution of the electron potential energy in silicon for $\Delta^{001}$ valley for a single quantum dot.

Figure 11 shows the enlarged fragment of Fig. 10. We display in all figures only negative values of the electron potential energy in silicon (below the bottom of the conduction band of non-strained silicon), because we are interested in the possible localization of electrons in the quantum dot. The germanium atoms and the silicon atoms forming Si-Ge bonds are not displayed too.

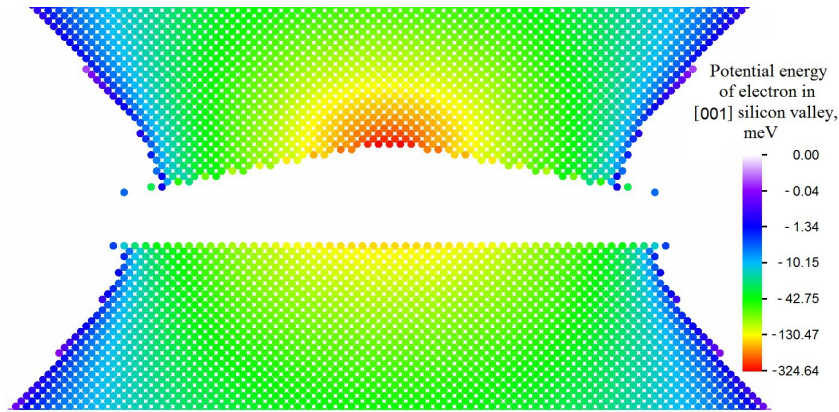

**Figure 11.** The enlarged fragment of Fig. 10. The distribution of the electron potential energy in silicon for $\Delta^{001}$ valley.

The distribution of the electron potential energy in silicon for $\Delta^{001}$ valley in the lateral section over the pyramid ($z$ = const) is presented in Fig. 12. The cross-section corresponds to the 4th layer of silicon atoms over the pyramid top (the 14th atomic layer in the numbering of Table 2).

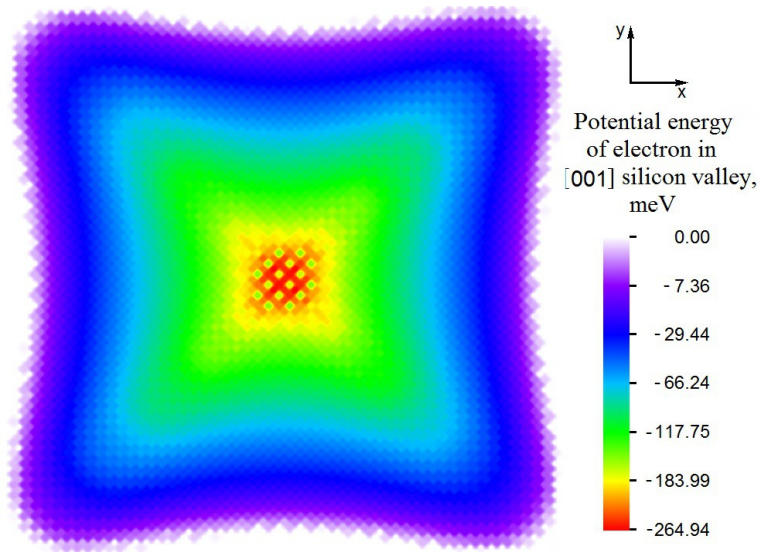

**Figure 12.** The distribution of the electron potential energy in silicon for $\Delta^{001}$ valley in the lateral section over the pyramid ($z$ = const, 14th atomic layer).

The distribution of the electron potential energy in silicon for $\Delta^{001}$ valley along the pyramid axis is presented in Fig. 13. It corresponds to Fig. 10-12. These values were obtained by averaging on cross-sections of 8 × 8 atomic layers of square column, also as well as the data in Table 2. The graph does not display higher energies for the germanium atoms and for the silicon atoms forming Si-Ge bonds. That is, the layers from -6 to 11 are not displayed.

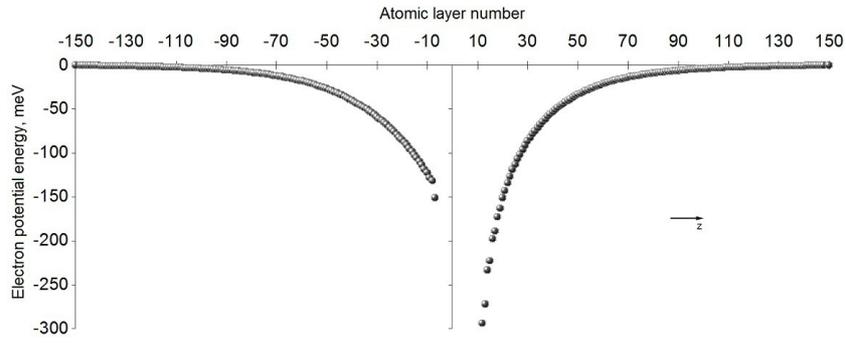

**Figure 13.** The distribution of the electron potential energy in silicon for $\Delta^{001}$ valley along the pyramid axis.

Figures 10, 11, and 13 show that there are two potential wells for electrons - above and below the pyramid. The potential wells are separated by a potential barrier formed by germanium atoms of the pyramid. Moreover, the potential well located above the pyramid apex is deeper than the potential well located under the pyramid.

## 2.4. The electron potential energy in silicon for $\Delta^{100}$ valley for the single pyramid

The distribution of the electron potential energy in silicon for $\Delta^{100}$ valley in the central section is displayed in Fig. 14. The enlarged fragment is presented in the right part of Fig. 13. In this case also, the potential wells are formed, which are about three times smaller than the potential wells for $\Delta^{001}$ valley. These wells in Fig. 14 are located in silicon above and below the wetting layer along the bottom perimeter of the pyramid.

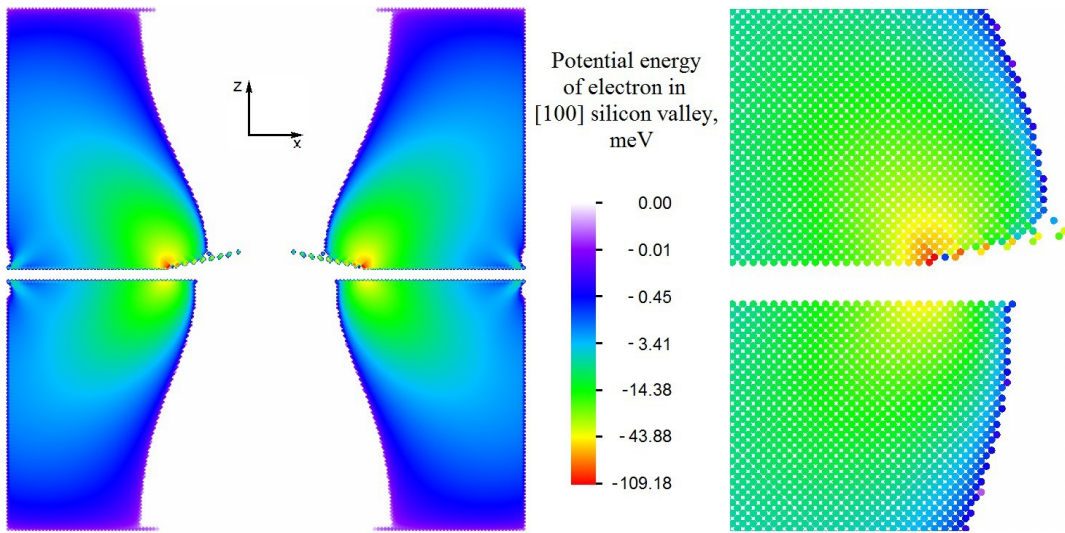

**Figure 14.** The distribution of the electron potential energy in silicon for $\Delta^{100}$ valley. On the right - the enlarged fragment.

The electron potential wells in silicon for $\Delta^{100}$ valley correspond to negative values of the components $u_{xx}$ of the strain tensor. It corresponds to a compression of silicon atoms along the *x* axis due to lateral repulsion of germanium atoms of the pyramid.

Figure 15 shows the distribution of the electron potential energy in silicon for $\Delta^{100}$ valley in the lateral section (*z* = const). This cross-section corresponds to the 4th layer of atoms over the wetting layer (3rd atomic layer in the numbering of Table 2). The electron potential well extends almost the entire length of the edges of the square base of the pyramid.

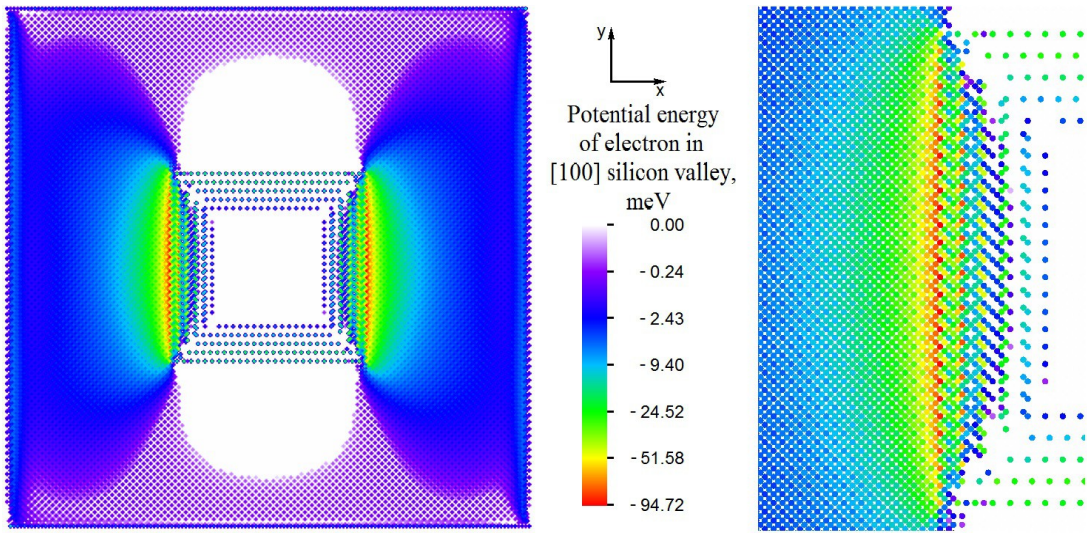

**Figure 15.** The distribution of the electron potential energy in silicon for $\Delta^{100}$ valley in the lateral section ($z$ = const, 3rd atomic layer). On the right - the enlarged fragment.

The potential wells in Fig. 15 are localized only near the two opposite edges of the base of the pyramid. Of course, there are the similar potential wells for the other two edges of the pyramid base, but they correspond to the other valley - $\Delta^{010}$.

Thus, there are 10 potential wells for electrons in silicon surrounding the germanium pyramid - 5 over the wetting layer, and 5 - under the wetting layer. Two potential wells for $\Delta^{001}$ valley are centered on the pyramid axis. Eight other smaller potential wells are localized around the base of the pyramid: 4 of them - for $\Delta^{100}$ valley, and 4 - for $\Delta^{010}$ valley.

### 3. The single pyramids of various sizes

The previous consideration dealt with the single pyramid with a base half-width of 55 atomic layers. Now we consider the dependence of the properties of a single pyramid on its size.

### 3.1. The strain distribution in germanium for the single pyramids of various sizes

Figure 16 presents in unified scale the distribution of the bulk density of the strain energy in germanium in the central section for the clusters containing the single pyramids with a base half-width of 140, 100, 60 and 20 atomic layers. The former height-to-base size ratio of 1 : 10 is kept for pyramids.

The largest pyramid in this Fig. 16 has a base half-width of 140 atomic layers. The total width of this pyramid is 38.0150 nm (between the centers of the extreme atoms).

Figure 17 presents the enlarged fragment of the distribution of the bulk density of the strain energy in germanium for the smallest pyramid from Fig. 16. The lower part of Fig. 17 shows all germanium atoms of the cluster in this cross-section. Some of the germanium atoms of the pyramid are not displayed on the top part of Fig. 17, because they have rather small strain energy for the used palette.

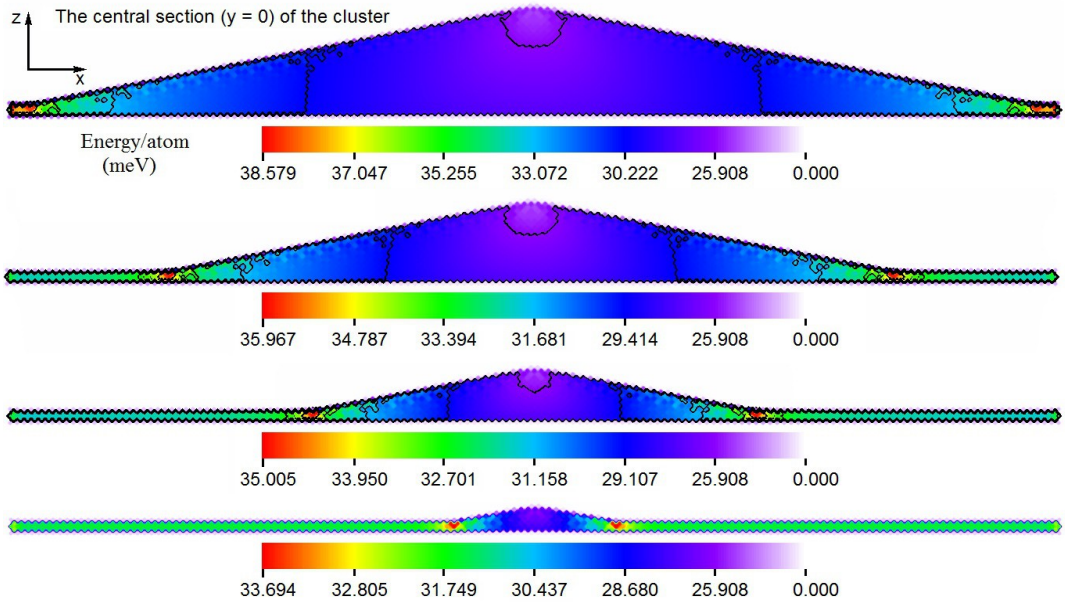

**Figure 16.** The distribution of the bulk density of the strain energy in germanium for the clusters containing the single pyramids with a base half-width of 140, 100, 60 and 20 atomic layers.

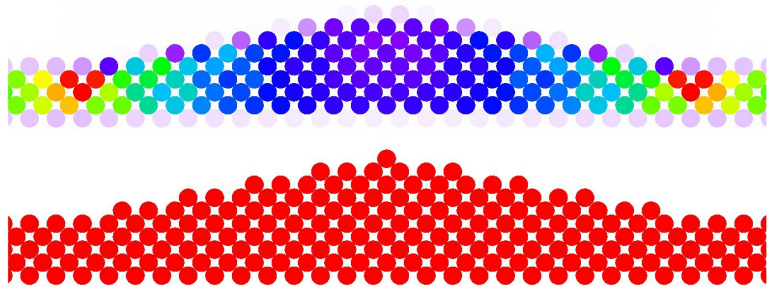

**Figure 17.** The enlarged fragment of Fig. 16 for the cluster containing the single pyramid with a base half-width of 20 atomic layers (the upper part). The lower palette from Fig. 16 is used. The lower part shows all germanium atoms of the cluster in this cross-section.

It is visible from Fig. 17 that the maximum strain energy is reached on the germanium atoms of the wetting layer. The numerical values of these energies are given in Table 3 and in Fig. 18. The atoms with maximum strain energy are located in the 2nd, 3rd and 4th coordination spheres in relation to the outermost atoms of the pyramid base. The similar picture is observed and for the pyramids of other sizes.

| Maximum strain energy density (meV/atom) | A base half-width of pyramid (atomic layers) | | | | | | | |
|---|---|---|---|---|---|---|---|---|
| | 20 | 40 | 55 | 60 | 80 | 100 | 120 | 140 |
| | 33.694 | 34.468 | 34.858 | 35.005 | 35.465 | 35.967 | 36.680 | 38.579 |

**Table 3.** The maximum strain energies of germanium atoms for clusters containing the pyramids of various sizes.

As it is seen from Fig. 18, the local deformation of the wetting layer near the edge of the pyramid increases with increasing the pyramid size. It corresponds to a compression of germanium atoms of the wetting layer along the $x$ axis due to lateral pressure of the increasing array of germanium atoms of the pyramid.

A rapid growth of the maximum strain energies for large pyramids can be explained by the influence of the adjacent pyramids, due to the almost periodic boundary conditions on the boundary of the cluster.

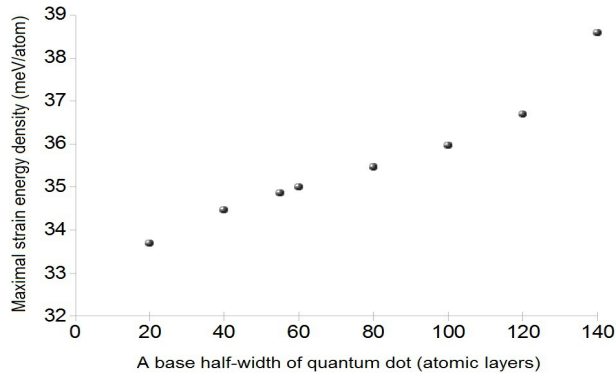

**Figure 18.** The maximum strain energies of germanium atoms for clusters containing the pyramids of various sizes.

### 3.2. The strain distribution in silicon for the single pyramids of various sizes

Now we consider the deformation in silicon surrounding the pyramid. Figure 19 shows the distribution of the bulk density of the strain energy in silicon in the central section for the pyramid with a base half-width of 140 atomic layers.

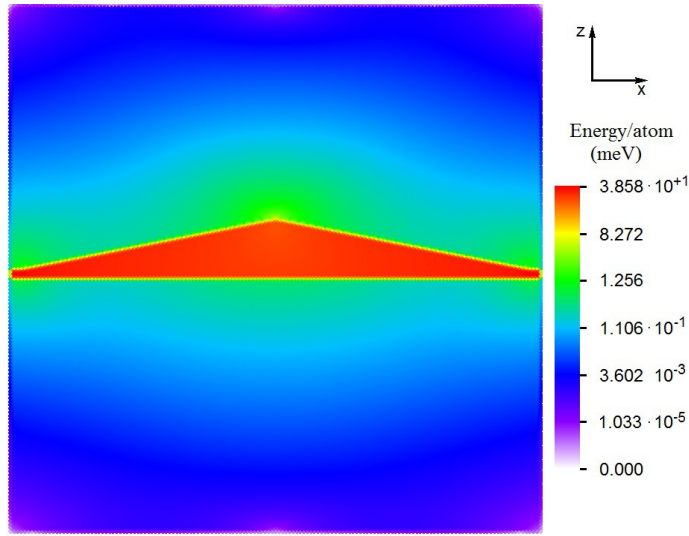

**Figure 19.** The distribution of the bulk density of the strain energy in silicon for the cluster containing a single quantum dot with a base half-width of 140 atomic layers.

Comparing Fig. 19, Fig. 7, and Fig. 1 from [1] shows a significant increase of the strain field with increasing the size of the pyramids. The main regions of deformation in silicon are localized near a pyramid axis. Smaller centers of deformation are localized at the edges of the pyramid base.

A noticeable increase in strain at the edges of the largest pyramid (Fig. 19) is largely due to the proximity of the border. The calculated distribution near the borders of the cluster corresponds approximately to the periodic conditions along the *x* direction. That is, the distribution, presented on Fig. 19, corresponds to a partial overlap of the strain fields of the similar neighboring pyramids with the distance between the bases of the pyramids in 20 atomic layers along the *x* axis.

### 3.3. The electron potential energy in silicon for $\Delta^{001}$ valley for the single pyramids of various sizes

Figure 20 shows the distribution of the electron potential energy in silicon for $\Delta^{001}$ valley in the central section of the pyramid with a base half-width of 140 atomic layers.

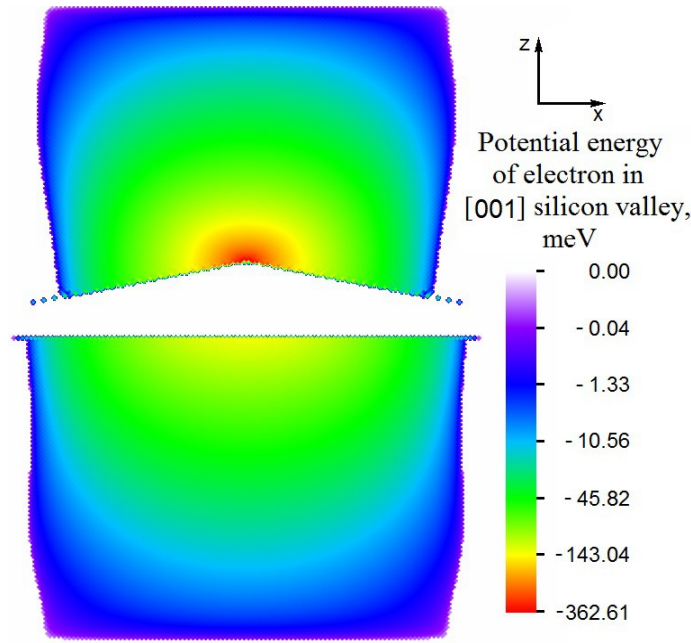

**Figure 20.** The distribution of the electron potential energy in silicon for $\Delta^{001}$ valley for the cluster containing a single quantum dot with a base half-width of 140 atomic layers.

Figure 21 shows the dependence on the size of the pyramids the values of the electron potential energy for $\Delta^{001}$ valley corresponding to the bottom of the potential wells, spatially localized above and below pyramids.

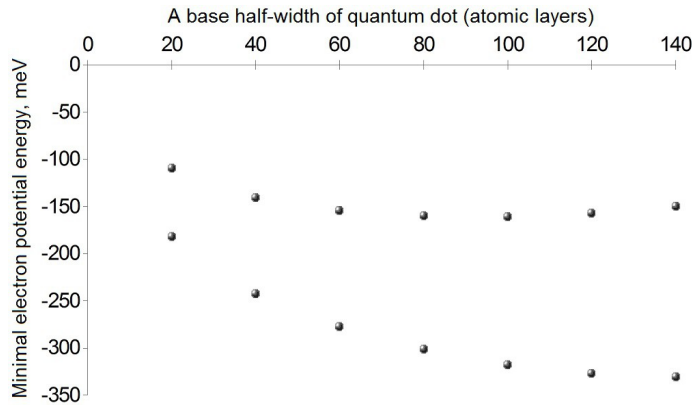

**Figure 21.** The values of the electron potential energy for the bottom of the potential wells for $\Delta^{001}$ valley for the single pyramids of various sizes. The lower set of points corresponds to the potential well above the pyramid apex. The top set of points corresponds to the potential well under the pyramid.

Numerical values for Fig. 21 are shown in Table 4.

| Minimum potential energy (meV) | A base half-width of pyramid (atomic layers) | | | | | | | |
|---|---|---|---|---|---|---|---|---|
|  | 20 | 40 | 55 | 60 | 80 | 100 | 120 | 140 |
| $E_1$ | -181.834 | -241.979 | -293.780 | -277.494 | -301.285 | -317.277 | -326.934 | -330.833 |
| $E_2$ | -109.281 | -140.687 | -151.064 | -154.143 | -159.986 | -160.863 | -157.481 | -149.938 |

**Table 4.** The values of the electron potential energy for the bottom of the potential wells for $\Delta^{001}$ valley for the single pyramids of various sizes. $E_1$ corresponds to the top potential well (above the pyramid apex). $E_2$ corresponds to the lower potential well (under the pyramid).

As can be seen in Fig.21, both potential wells for small pyramids deepen with increase in the sizes of pyramids. For large pyramids energy decrease practically stops for the top potential hole and even is replaced with increase in energy for the lower potential hole. This effect is also connected with the almost periodic boundary conditions on the boundary of the cluster, i.e., with the influence of the adjacent pyramids.

### 3.4. The electron potential energy in silicon for $\Delta^{100}$ valley for the single pyramids of various sizes

Figure 22 shows the distribution of the electron potential energy in silicon for $\Delta^{100}$ valley in the central section for the pyramid with a base half-width of 140 atomic layers. Here is also observed an increase in the depth of the potential well in comparison with the smaller pyramids (Fig. 23).

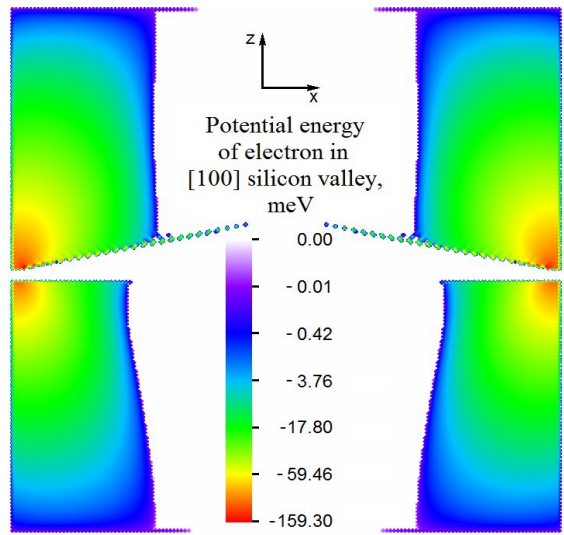

**Figure 22.** The distribution of the electron potential energy in silicon for $\Delta^{100}$ valley for the cluster containing the pyramid with a base half-width of 140 atomic layers.

More sharp deepening of the potential wells for large pyramids in Fig. 23 is again connected with the almost periodic boundary conditions on the boundary of the cluster, i.e., with the influence of the adjacent pyramids.

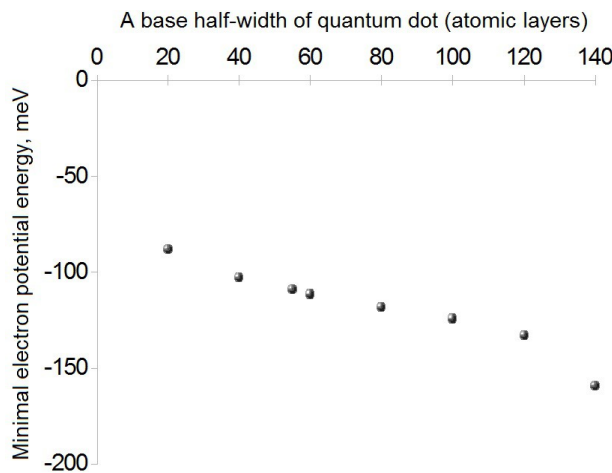

**Figure 23.** The values of the electron potential energy for the bottom of the potential wells (above the wetting layer) for $\Delta^{100}$ valley for the single pyramids of various sizes.

Numerical values for Fig. 23 are shown in Table 5.

| Minimum potential energy (meV) | A base half-width of pyramid (atomic layers) | | | | | | | |
|---|---|---|---|---|---|---|---|---|
| | 20 | 40 | 55 | 60 | 80 | 100 | 120 | 140 |
| | -88.18 | -102.97 | -109.18 | -111.43 | -117.95 | -124.34 | -133.03 | -159.30 |

**Table 5.** The values of the electron potential energy for the bottom of the potential wells for $\Delta^{100}$ valley for the single pyramids of various sizes.

## IV. CONCLUSIONS

The detailed modeling of the elastic strain fields near the single quantum dots in the Ge/Si system has been carried out. This modeling expands some results presented in [1]. The cluster approximation and the discrete atomistic description of interatomic interactions within the Keating model was used.

The calculation results confirm the results obtained earlier in the discrete-continuous model in [9] and in works of other authors.

The atomistic approach is more adequate to describe the properties of quantum dots on the nanometer scale in comparison with the continuum approach, since it does not contain a rough continuum approximation of the originally discrete atomic task.

In our calculations, we solve the computational problems that arise when using a purely atomistic approach.

It is shown that the strain field in silicon decreases sufficiently rapidly with distance from the center of the quantum dot, so the influence of the cluster boundary is observed only for very large quantum dots. Therefore, use of large clusters containing atoms of 150 coordination shells, together with suitable boundary conditions, allows us to calculate accurately the deformation field for the quantum dots with dimensions of about 20 nm. Even for large quantum dots with dimensions close to the dimensions of calculated clusters, it is possible to interpret the influence of the cluster boundary as the influence of neighboring quantum dots, because the used boundary conditions correspond approximately to the periodic conditions at the borders.

The calculations show that there are 10 regions of increased strain, spatially localised in silicon, in which formation of the localized quantum states for electrons are possible. Strain-induced electron localization is not so strong and for its enhancement the multilayered quantum dots structures were proposed and realized [13]. This problem was considered in our previous work [1]. In [14] the electron localization in different Δ-valleys was demonstrated for structures with double quantum dots layers. Simultaneous localization of electrons at the quantum dots apexes and at the base edges was observed in [15].

The problem of the efficiency of electron localization in $\Delta^{100}$ and $\Delta^{001}$ valleys requires further investigation, including quantum-mechanical calculations.

## ACKNOWLEDGMENTS

This study was supported within the Integration project of the Siberian Branch of the Russian Academy of Sciences no. 43 "Development of the Physical Principles of Logic Gates Based on Quantum Dot Nanostructures".